\documentclass{article}

 \usepackage[preprint,nonatbib]{neurips_2026}

\usepackage[utf8]{inputenc} 
\usepackage[T1]{fontenc}    
\usepackage{hyperref}       
\usepackage{url}            
\usepackage{booktabs}       
\usepackage{amsfonts}       
\usepackage{nicefrac}       
\usepackage{microtype}      
\usepackage{xcolor}         
\usepackage{subcaption}
\usepackage{amsmath}
\usepackage{tcolorbox}
\usepackage[capitalize,noabbrev]{cleveref}
\usepackage{enumitem}
\usepackage[colorinlistoftodos]{todonotes}
\setlist[enumerate]{leftmargin=4ex, nosep, noitemsep}
\setlist[itemize]{leftmargin=4ex, nosep, noitemsep}

 \usepackage[style=numeric, natbib,sorting=none]{biblatex}
\title{EASE Configuration Facilitates A Reproducible Science of LLM Social Simulations}

\author{%
  Sneheel Sarangi\textsuperscript{1,2}\thanks{Equal Contribution.} \And
  Maximilian Puelma Touzel\textsuperscript{1,3}\footnotemark[1] \And
  Aur\'{e}lien B\"{u}ck-Kaeffer\textsuperscript{1,2,4} \AND
  Zachary Yang\textsuperscript{1,2,4} \And
  Jean-Fran\c{c}ois Godbout\textsuperscript{2,3} \And
  Reihaneh Rabbany\textsuperscript{1,2} \AND
  \textnormal{\textsuperscript{1}McGill University, \textsuperscript{2}Mila - Quebec Artificial Intelligence Institute,} \\
  \textnormal{\textsuperscript{3}Universit\'{e} de Montr\'{e}al, \textsuperscript{4}Ubisoft La Forge}
}

\begin{document}

\maketitle

\begin{abstract}
LLMs are increasingly deployed to simulate social interactions, yet many of the existing simulators remain \textit{ad hoc} and monolithic. This lack of architectural standardization prevents reproducible research and complicates downstream evaluation. We advance a rigorous science of LLM-based multi-agent simulation by modularizing core components into Environments, Agents, Simulation engines, and Evaluation metrics (EASE). We demonstrate the utility of EASE configuration by wrapping it in an experimental study schema for orchestrating workflows centered around answering explicit research questions in generated scenarios. We contribute \texttt{SiliSocS}, an open-source, research-ready Silicon Society Sandbox implementing study-structured EASE configuration to enable highly configurable and reproducible LLM-based social simulations. Using \texttt{SiliSocS} and EASE, we present three case studies, showcasing the system's comprehensive assessment of existing questions, ability to dive deeper into complex questions, and elaboration of existing studies, respectively. Together, these case studies highlight the limitations of current modeling approaches and isolate the impacts of design choices on key results. 
\end{abstract}

\section{Introduction}


Large Language Models (LLMs) have opened the possibility of simulating social systems exhibiting the complexity of human language: human societies as well as systems of interacting, goal-directed AI agents. 
Unlike earlier Agent-Based Models, often criticized as simplistic \citep{xi2023risepotentiallargelanguage}, LLMs can leverage vast cultural knowledge from pretraining and can be fine-tuned within agent harnesses designed either to solve generic tasks or to more closely mimic the social behavior of real humans captured in text. 
Researchers across computational social science, public health, and policy analysis are now building LLM-based simulations to study a wide variety of phenomena, such as opinion dynamics \citep{neumann2025usellmssimulateopinions,zuo2025mtosllmdrivenmultitopicopinion}, cooperation \citep{Piatti2024, smith2024the}, and information propagation \citep{yalabadi2024controllingmisinformationdiffusionsocial, orlando2025emergentcoordinatedbehaviorsnetworked}. Engineers and AI safety researchers are similarly modelling how agents interact with one another in multi-turn settings \citep{shapira2026agentschaos,vallinder2024culturalevolutioncooperationllm}.

However, this rapid proliferation has outpaced the field's capacity to validate what these simulations produce. Evaluation practices are underdeveloped, fragmented, and rarely shared \citep{anonymous2026position}. Simulators are engineered to scale up to a million agents \citep{yang2024oasis}, yet remain difficult to use for scientific inference because their alignment with real-world behavior is weakly evidenced or unknown \citep{larooij2025largelanguagemodelssolve}. Furthermore, because there is no consensus on architecture, behaviors observed within these environments risk being restrictive artifacts of the specific simulator's design rather than generalizable phenomena. Ad-hoc environment parameters and prompt framing can be inadvertently tuned to guarantee desired social outcomes or surface-level mimicry \citep{zhou2026pimmur}.


These failures trace to a common architectural problem: existing simulators are monolithic systems in which the contributions of individual components are custom-built and deeply entangled. When a simulation produces an interesting outcome, it is typically very difficult to determine whether it reflects genuine emergence, training-data bias, prompt engineering, or the interaction protocol. This prevents reproducible research and principled knowledge accumulation.

Similar to past technological design innovations linking objects and applications (\textit{e.g.} the internet protocol), we argue the field needs a science of simulations built on design protocols: with simulators that can lead to rigorous, reproducible, and hypothesis-driven science. A simulator's scientific value lies not in feature richness, but in causal transparency: the ability to ask, for any observed phenomenon, which component is responsible?

To address this entanglement, we present EASE, a modular framework for LLM social simulation, named for the four design spaces it separates: environment, agent, simulation engine, and evaluation design. This modularity allows researchers to intervene on a single component while holding the rest of the system constant. Crucially, EASE includes the evaluation space, ensuring that metrics, structural safeguards, and behavioral probes are treated as explicit, reproducible components rather than just \textit{ad hoc} post-processing. To demonstrate the utility of EASE configuration, we introduce an experimental methodology that embeds individual simulation runs into a structured study design, explicitly defining what is being tested, what is held fixed, and what constitutes valid evidence.

By adopting EASE as an explicit architectural paradigm, vague inquiries such as ``Do LLMs exhibit emergent cooperation?'' are translated into precise, scientifically testable hypotheses: ``Does cooperation, as defined in the collective animal behavior literature, persist when we alter the agents' memory architecture while holding the environment's interaction protocol and the simulation engine fixed?'' Ultimately, EASE configuration equips researchers with the structural scaffolding necessary to conduct controlled ablation studies, rigorous sensitivity analyses, and systematic cross-model comparisons—providing the essential experimental toolkit required to move the field from observational mimicry to robust causal inference.



\textbf{Our contributions (\Cref{fig:schematic}):}

\begin{enumerate}[label=C\theenumi]

\item \textit{A study design methodology} for hypothesis-based iterative experimentation. 
\item \textit{EASE simulator design and configuration schema}, outlining key building blocks of a Generative Agent-Based Simulator as independent components and subcomponents.


\item \textit{An open-source codebase}, \texttt{SiliSocS}\footnote{https://github.com/sandbox-social/silisocs} (Silicon Society Sandbox) that implements the EASE framework and the proposed study structure, and is thus structured for frictionless reproducibility.


\item Three \textit{case studies} on style diversity, engagement behavior, and echo chamber formation to illustrate distinct use cases.

\end{enumerate}

\section{Related Work}

\paragraph{Social simulation with LLMs.}
The use of LLMs to simulate social dynamics has grown rapidly, spanning opinion formation \citep{neumann2025usellmssimulateopinions}, collective decision-making \cite{ferrarotti2026generativeaicollectivebehavior}, and online discourse \cite{ng2025llm}. Early work demonstrated that LLM-based agents can reproduce stylized facts from classic social science models, prompting enthusiasm about their potential as \emph{in silico} laboratories for studying human behavior. However, this enthusiasm has been tempered by a growing body of methodological critique. \citet{zhou2025pimmur} and \citet{barrie2025emergentllmbehaviorsobservationally} question whether behaviors observed in multi-agent simulations genuinely \emph{emerge} from agent interaction or are instead artifacts of the underlying model's training distribution. Thus, raising concerns about the inferential validity of results derived from such systems. \citet{larooij2025fixsocialmediatesting} level a parallel critique at simulations of social media platforms, arguing that without disciplined testing protocols, conclusions about platform dynamics remain unfalsifiable. \citet{anonymous2026position} point out the validation gap that exists in current social-simulation systems, arguing for the need for shared evaluation protocols. These criticisms motivate the central design goal of EASE: making every methodological choice explicit, swappable, and individually auditable.

\paragraph{Simulation frameworks.}
Several recent projects have pursued more comprehensive infrastructure for LLM-driven social simulation. AgentSociety \citep{piao2025agentsocietylargescalesimulationllmdriven} and OASIS \citep{yang2024oasis} each provide large-scale environments with rich agent populations and interaction channels. While these systems demonstrate impressive scale, they tend toward monolithic architectures in which agent design, environment logic, and evaluation are tightly coupled, making it difficult to isolate the effect of any single design decision or to port experimental protocols across settings. EASE addresses this gap by decomposing the simulation pipeline into independently configurable modules that our study schema wraps into an experiment workflow that foregrounds hypothesis specification and cross-scenario comparison.

\paragraph{Design philosophy and orchestration.}
Our modular approach builds most directly on the design philosophy of Concordia \citep{vezhnevets2023generativeagentbasedmodelingactions,vezhnevets2025multiactor,leibo2026theoryappropriatenessaccountsnorms}, which introduces a library of reusable orchestration objects (i.e., game masters, components, and action spaces) for constructing applications such as generative agent-based simulations. However, Concordia does not prescribe how to organize those simulations into controlled experiments or how to systematically vary conditions for reproducibility. We address this by adopting the same ``meeting-in-the-middle'' design \citep{design} one level higher in abstraction to modularize research workflows themselves into composable building blocks. EASE treats a simulator (potentially built with Concordia-style primitives) as one element within a broader experiment object that additionally specifies research questions and evaluation pipelines. 

\paragraph{Connections to social science.}
A key audience for principled simulation infrastructure is the social science community, where LLM-based modeling is increasingly seen as a tool for theory development and counterfactual reasoning. \citet{ding2025understandingworldpredictingfuture} and \citet{zhou2025socialworldmodels} position LLM simulations as ``social world models'' capable of capturing complex societal dynamics, while \citet{park2022socialsimulacracreatingpopulated} demonstrates the generation of synthetic populations for studying social phenomena. These efforts illustrate genuine demand for simulation-based inquiry in the social sciences, yet each work constructs its own bespoke pipeline, limiting reproducibility and cross-study comparison. Our \texttt{SiliSocS} codebase implementing EASE and our study schema is designed to lower the barrier for social scientists by providing a standardized, configurable entry point: researchers specify hypotheses and scenarios declaratively, and the framework handles orchestration, execution, and structured output for downstream analysis.

\section{Methodology}

\subsection{EASE-based Simulation Framework: Components of a Simulation}
\begin{figure}[t!]
    \includegraphics[width=.98\textwidth]{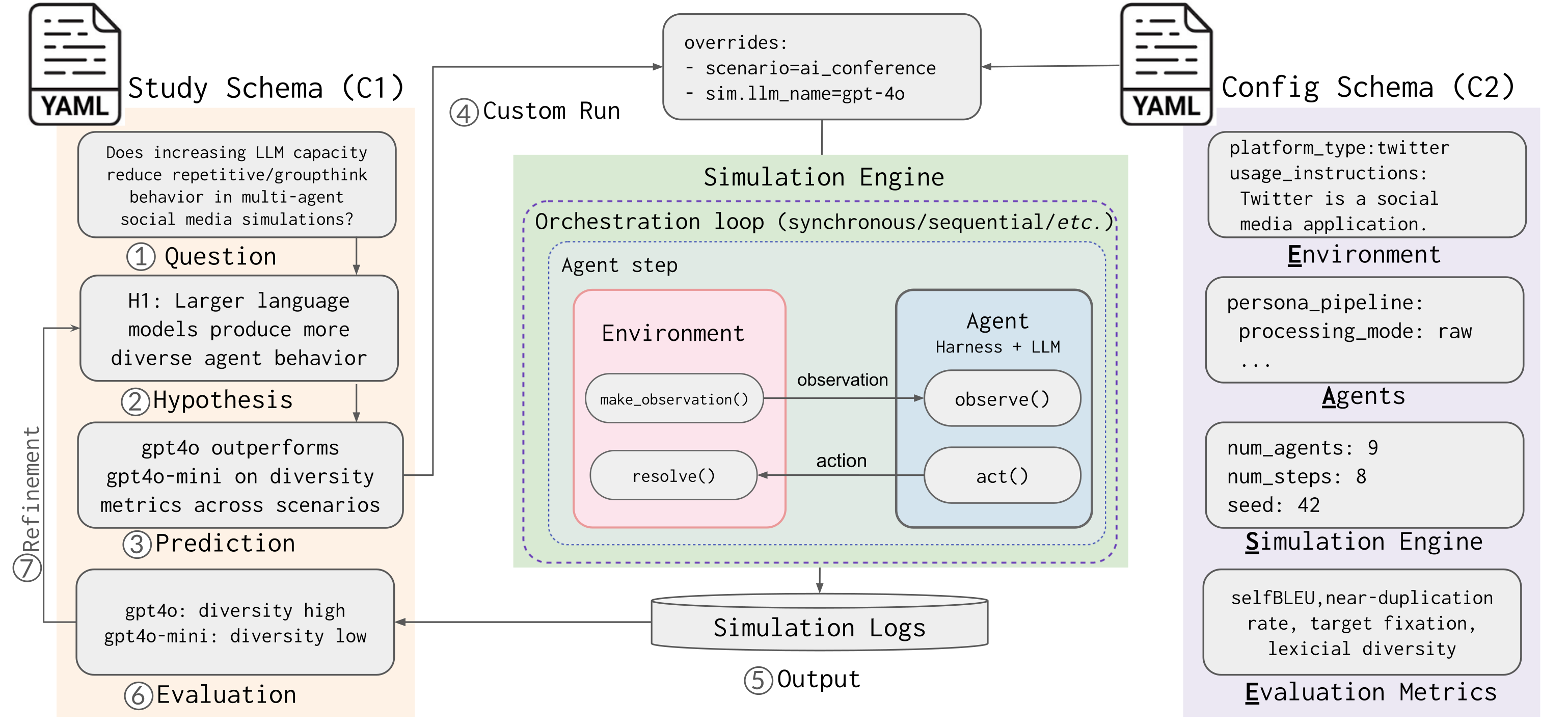}
\caption{\textit{System Design Exemplified with Application To Style Diversity}. The framework consists of EASE simulation configuration (C2; right): Environments, Agents, Simulation Engine, and Evaluation Metrics. These are used to configure a (\textit{e.g.}, Concordia) simulation engine (middle) to run custom simulations within a 7-step research cycle structured in our proposed study schema (C1; left). The entire system is provided in an open-source codebase, \texttt{SiliSocS} (C3).
}
\label{fig:schematic}
\end{figure}

Social simulation design requires coordinating several complex, interacting elements. To create an effective framework for understanding and decomposing social simulations, we conceptualize the architecture through both theoretical and practical lenses. We classify social simulation design into four core components: Environments, Agents, Simulation engine, and Evaluation metrics (EASE) (\Cref{fig:schematic}(right)). Since the scientific value of simulation results comes from how they are interpreted, we strongly advocate for including evaluation metrics in the configuration. Below, we discuss each component in detail alongside the core subcomponents we decouple. We note that we are not advocating for every subcomponent to be a strictly distinct software module; in practice, roles may be combined or delegated. However, by establishing this modular framing, the responsibilities of each component become transparent, allowing for clean variable isolation during experimentation.

\paragraph{Environment} The environment constitutes the social space within which agents interact. Practically, we can think of it as two categories of subcomponents: those related to the backend, and those with the agent-environment interface  While subcomponents for the environment backend are largely scenario-dependent (\textit{e.g.}, a recommendation algorithm is critical for a social media simulation), we explicitly isolate the environment-agent interaction layer in an interfacing object inspired by the game master abstraction proposed in the Concordia framework. This bridging layer primarily resolves two questions:
\begin{enumerate}
\item \textit{Observation Formation}: How is the global state of the environment filtered and curated for an individual agent's consumption?
\item \textit{Action Resolution}: How are an agent's intended actions translated and executed within the environment's rules set?
\end{enumerate}

Reproducibility demands specifying the interfacing operations explicitly. The scope of observability given to agents can have a strong impact on emergent agent behavior \cite{ferrarotti2026generativeaicollectivebehavior, orlando2025emergentcoordinatedbehaviorsnetworked}. Furthermore, decoupling an agent's intended action from the environment's permitted actions reduces structural ambiguity, ensuring a clean boundary between agent decision-making and environmental physics.

\paragraph{Agents} Agents are the individual modeling units of the simulation. Their behavior, both in aggregate and distribution, generates the primary insights desired from the simulation task. In LLM-based simulations, agent design combines an LLM oracle with an agent harness divided into the subcomponents that collectively govern perception, cognition, and action:
\begin{enumerate}
\item \textit{Model}: The LLM oracle employed by the agent harness (\textit{e.g.}, answering agent planning queries), the choice of which can strongly influence the agent's reasoning complexity and general behavioral priors.
\item \textit{Instructions}: Static prompts and constraints providing relevant scenario context and behavioral boundaries.
\item \textit{Persona}: The set of modules responsible for grounding the simulated agent in a realistic, consistent character. Personas can be prompt-based \citep{venkit2026needsociallygroundedpersonaframework} or model-based, such as via trained LoRAs \citep{buckkaeffer2025blueprint}.
\item \textit{Memory Structure}: The storage and retrieval mechanism for agent experiences. This is especially critical for social simulations because long-horizon interactions risk saturating context windows, and performance—particularly in the smaller LMs often used to scale simulations—degrades as context lengthens.
\item \textit{Cognitive Architecture}: The internal reasoning framework that dictates how an agent processes memory and context before taking action.
\end{enumerate}

\paragraph{Simulation Engine} While some \textit{ad hoc} system designs may couple the engine directly with the environment, we believe that a structured science of simulations should treat the engine as a distinct entity. The Simulation Engine acts as the orchestrator: it enforces the temporal design of the simulation, discretizes time into manageable episodes, and dictates how agents act within those bounds (\Cref{fig:schematic}(center)). Its core components include:
\begin{enumerate}
\item \textit{Inter-Agent Dynamics}: Determines the concurrency of the simulation (e.g., whether agents take actions sequentially, synchronously, or in parallel).
\item \textit{Agent Action-Making}: Defines the constraints of an agent's turn within a single episode (e.g., action limits or rate-limiting).
\item \textit{Agent Activity}: The selection mechanism that determines which subset of agents is active during any given episode.
\end{enumerate}

\paragraph{Evaluations} The evaluation layer standardizes how simulation outcomes are measured and validated. It can be broadly divided into three subcomponents:
\begin{enumerate}
\item \textit{Reference Scenario}: In general, a scenario can be defined as any set of components held fixed over a series of experiments to test a hypothesis. In practice, this can involve semantic content serving as the social context in which agents act, environment structure, etc. Often, simulations draw this context directly from real-world datasets to enable comparative validation. A diversity of scenarios provides a test of reliability over contextual factors.
\item \textit{In-Simulation Probes}: Active measurement tools deployed during runtime, such as structured surveys or queries posed to agents between episodes to track latent state changes.
\item \textit{Post-Hoc Analysis}: The suite of metrics and statistical analyses applied to the simulation's final output logs to quantify macroscopic phenomena.
\end{enumerate}

\subsection{Experimental Study Structure}

The simulator design space grows combinatorially: a study may vary agent harness, base LLM, persona construction, memory policy, observation formation, action resolution, timeline algorithms, scenarios, evaluation metrics, random seeds, and more. Exhaustive sweeps are infeasible, as LLM simulations require many model calls over long contexts and stochastic variance demands repeated runs per condition. EASE supplies the component vocabulary to map this space. We introduce a study schema that helps navigate by recording the path taken through it.

\paragraph{Hypothesis-driven exploration.} Discovery-based sweeps can surface unexpected regularities but are resource-intensive \cite{larooij2025largelanguagemodelssolve}. Given the cost of LLM simulations, we advocate a hypothesis-driven primary mode: a research question is translated into concrete hypotheses about what controls a phenomenon, predictions are derived, and evaluations are selected that can falsify those predictions. This workflow is also well-suited to agent-human collaboration, being tedious yet highly templated.

\paragraph{The study schema.} We encode this workflow as a structured text schema. \Cref{fig:schematic}(left) that makes each step explicit: (1) question, (2) hypotheses, (3) predictions, (4) runs, (5) outputs, (6) evaluation, and (7) refinement into follow-up hypotheses. Each hypothesis specifies its independent variable, the EASE components held fixed, and the metrics that constitute valid evidence. Logging artifacts against the hypothesis they served makes sequential exploration auditable: a researcher can run the minimal comparison needed, inspect the result, and branch to the next most informative question while preserving a trail of what was tested, what was held fixed, and why. This adapts established traditions of screening, sequential, and iterative designs from sensitivity analysis to a setting where consequential factors are modular components rather than numeric parameters.

\paragraph{Scenarios and initialization.} Hypothesis testing requires a controlled starting point. A scenario object bundles settings, actors, and context into a reusable template. Initialization binds scenario agents and environment by specifying initial state, persona assignments, interaction topology, memory seeds, initial content streams, random seeds, and any pre-simulation history. Different initializations can yield different downstream behavior, even with all other components fixed, so the schema records initialization explicitly.

\paragraph{Reference scenario generation.} Reproducibility in social simulation is hampered by limited access to real-world reference data. We provide an LLM-driven scenario generation pipeline: a scenario schema, and coding-agent skills that walk users through specifying instances for their question; a companion skill generates scenario sets spanning the range of signals a simulation might produce about a question, supporting tests of reliability across contextual (\textit{e.g.} semantic) factors. As a best practice, evaluation metrics should be kept out of the generation process to avoid correlating scenarios with metrics. Scenarios used in our case studies are described in Appendix  \Cref{app:refscen}.

\subsection{The \texttt{SiliSocS} Codebase}

\texttt{SiliSocS} implements the EASE configuration framework and the experimental study wrapper (with scenario generation) as a research-oriented simulation sandbox. The codebase is designed around two entry points. The first is a \textit{configuration entrypoint}: researchers can straightforwardly compose and run simulations from existing agents, environments, engines, scenarios, and evaluations using declarative configuration files. This supports reproducible experimentation because design choices are expressed as explicit configuration objects rather than hidden inside experiment scripts.
The second is a \textit{design entrypoint}: researchers can extend the framework by implementing new components. New agent architectures, memory modules, environment backends, game-master interfaces, activity schedules, action-resolution rules, probes, and \textit{post hoc} evaluations can be added as framework elements and then exposed through the same configuration interface. This lets users move between using \texttt{SiliSocS} as an experiment runner and using it as a library for designing new simulation components, or replicating existing ones.


Each run produces structured artifacts, including the resolved configuration, random seeds, scenario specification, event logs, agent actions, in-simulation probe outputs where applicable, and post hoc evaluation files. Study files link these artifacts back to hypotheses and conditions, so that a result can be traced from a claim in the paper to the configuration and logs that produced it. Studies and scenarios follow schemas that are used by a structured generation process. The process is highly documented and can be followed in the command line at various levels of user input, from a single, high-level description to a detailed sequential specification. We ask users to review all configurations not explicitly entered. 
We discuss further code details in \Cref{app:codebase}.

\section{Case Studies} 
\begin{figure}[t!]
    \includegraphics[width=\textwidth]{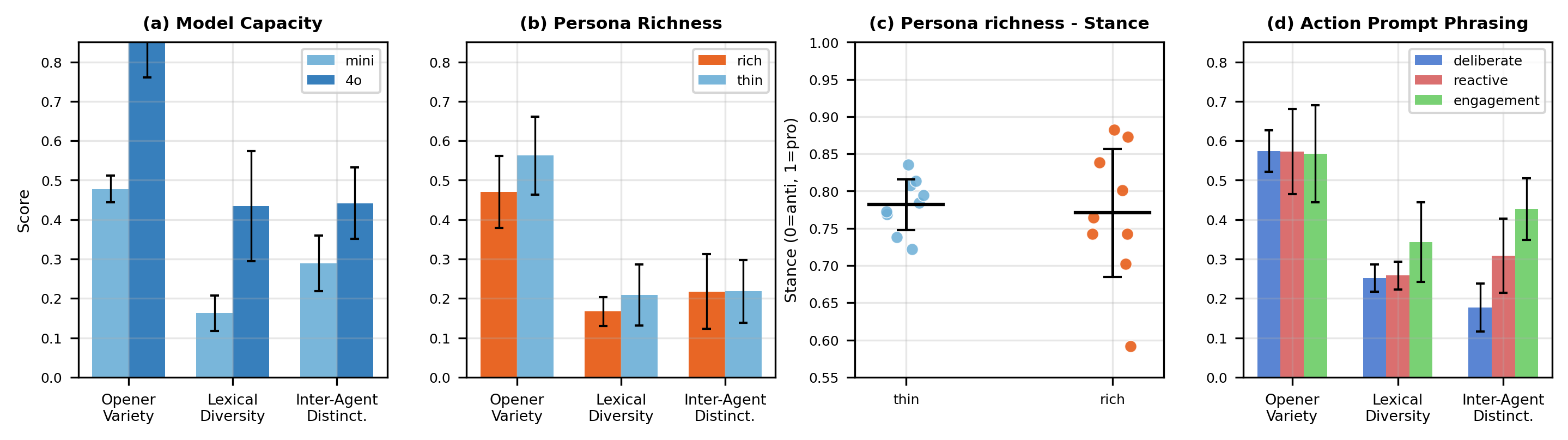}
\caption{\textit{Style diversity study results.} (a) \texttt{gpt4o} outperforms \texttt{gpt4o-mini} in having more diverse responses. (b) Post diversity seems unaffected by stronger grounding of agents using rich personas (\texttt{gpt4o-mini} was fixed here). (c) Posts are, nevertheless, more diverse with rich personas, just in stance, not in lexical diversity. (d) Action Prompt rephrasing for distinct goals gives has little effect on within agent diversity, but gives a range of inter-agent diversity even larger than that given by model strength (see panel (a)).}
\label{fig:style}
\end{figure}
\subsection{Case Study 1 — Style Diversity: comprehensive assessment of an existing question.}
Many works have identified stylistic diversity as important for face validity \citep{anthis2025position}, both in one-shot survey settings and in multi-turn multi-agent interaction settings. We undertook a study to reveal sources of stylistic diversity (the study file is in the Appendix \Cref{sec:studies}). As metrics, we chose standard metrics of lexical diversity comparing text pairs from both within and across agents (see Appendix \Cref{sec:div_metrics} for definitions and explanations). For method details, see Appendix \Cref{sec:diversity_details}.

Face validity has improved significantly over model versions. To see if style diversity specifically remains an issue, we can evaluate diversity metrics over model strength to see trends (in which case the problem is unlikely to persist as models continue to improve). We thus made H1 about model strength (see \Cref{fig:style}a), showing that, indeed, \texttt{gpt4o} achieved much more diverse style in posts than \texttt{gpt4o-mini}. How significant is this difference relative to other simulation components? We selected and worked through alternative candidates using EASE configuration modifications. First, we looked at model temperature (H2), finding largely null results. Next, we considered richer persona definitions (H3). Surprisingly, the latter also appeared to give little difference (see \Cref{fig:style}b). To see exactly where the effects of richer personas were being washed out, we utilized the probe functionality of the simulator to collect reflections and opinions and indeed found that rich personas do exhibit more diverse survey responses (see Appendix \Cref{fig:stance}). 
That said, agents, and especially those with rich personas, can have different stances when talking about the same topic, and indeed we find a larger spread of stances in posts (scores from \texttt{cross-encoder/nli-deberta-v3-small}) in the rich persona condition relative to the thin one (see \Cref{fig:style}c). The discrepancy is a cautionary tale about the importance of the chosen metric in surfacing a concept of interest (here, diversity).
Finally, we questioned how much diversity can be obtained using a weak model and simply varying the action prompt. We designed three action choice instruction prompts derived from different goals: from deliberate (excerpt: `encourages careful, persona-faithful reasoning before acting'), to reactive (excerpt: `encourages impulsive, immediate response to content'), to engagement (excerpt: `encourages reach-maximising, audience-seeking behaviour') (H4; \Cref{fig:style}d). We found that inter-agent distinctiveness increased significantly across call-to-action prompt goals. The range of distinctiveness even exceeds the difference observed between the weak and strong models from H1, suggesting that prompt engineering can close the gap when using a weaker model.

\subsection{Case Study 2 — Engagement Dynamics: diving deeper into a complex question via iterative branching.}
\label{subsec:case-study-engagement-dynamics}

\begin{figure}[t!]
\centering
\includegraphics[width=\textwidth]{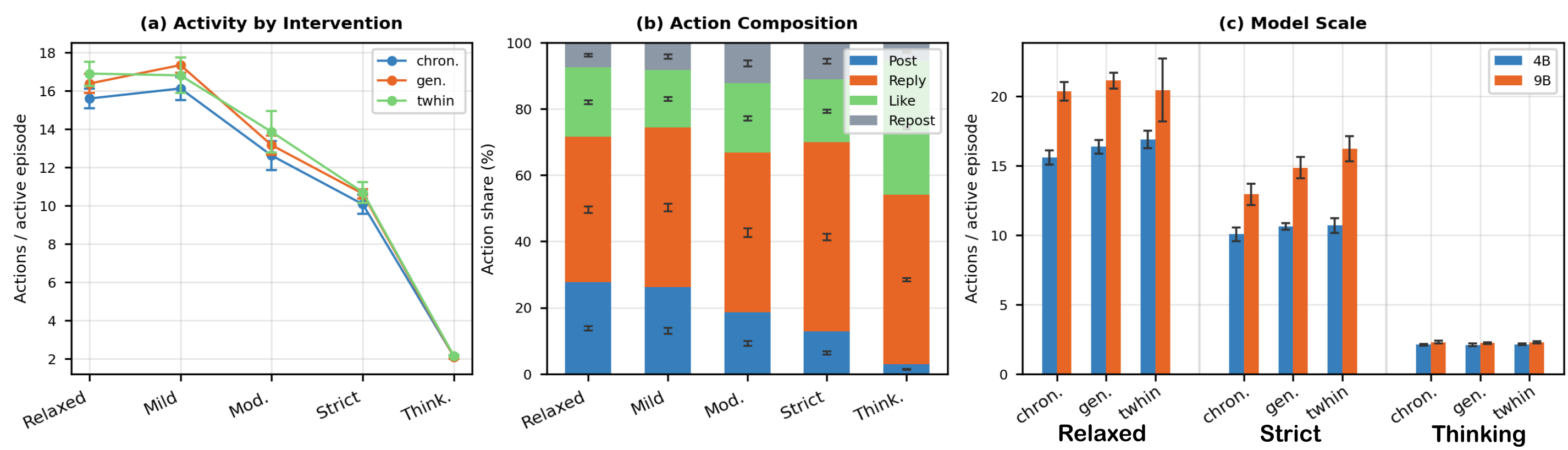}
\caption{Engagement case study panels: (a; left) total actions per active agent per active episode for the Qwen3.5-4B model (4b), (b; middle) action composition (shares across channels) across prompt interventions for 4b, (c; right) total actions across recommendation algorithms and intervention strength, comparing model scales (Qwen3.5-9B (9b) vs 4b). Error bars denote 95\% confidence intervals across seeds. }
\label{fig:engagement-case-study}
\end{figure}

Our second case study examines engagement dynamics in a social-media-like simulation. The study asks whether algorithmically curated timelines elicit more agent activity than chronological timelines, and then uses the study schema to identify which simulator components contribute to this effect. We use the number of actions as a metric of engagement, while also tracking action composition. Since most LLM-based social simulations fix the number of actions agents may take per timestep \citep{bückkaeffer2026textitsiliconsocietycookbookdesign, yang2024oasis, coppolillo2025engagementdrivencontentgenerationlarge}, this case study treats the action budget itself as an experimental component.

We fix the scenario, agent personas, background context, and timeline size at 10 posts per timestep, and compare a follower-chronological baseline with two recommendation systems: a general embedding-based recommender and a TwHIN-based recommender \citep{El_Kishky_2022}, implemented following OASIS \citep{yang2024oasis}. We primarily use Qwen3.5-4B \citep{qwen3.5}; full settings, prompts, seed-matched statistics, and $p$-values are provided in \Cref{app:engagement_dynamics}.

The study began with the hypothesis that curated timelines should increase engagement relative to chronological timelines. Under an initial cap of 12 actions per active agent per episode, however, total activity showed no meaningful differences across timeline algorithms. Because all conditions approached the action cap, the indistinguishability was more plausibly due to a simulation-engine ceiling than to evidence that timeline curation has no effect.

Relaxing the action budget from 12 to 20 revealed the hidden separation: the TwHIN recommender produced significantly more activity than the chronological timeline, reaching 16.91 actions per active agent per active episode compared with 15.60 for chronological (\Cref{tab:engagement-ladder}; \Cref{fig:engagement-case-study}a). Thus, the original timeline hypothesis was supported, but partially masked by the simulation engine.

This result also exposed a realism problem. With only 10 timeline posts per episode, more than 16 actions per active agent is implausibly high. We therefore branched to agent-level prompt interventions that encouraged more realistic behavior, with the strongest condition asking agents to reason before selecting actions; the full prompts are in \Cref{app:intervention_prompts}. These interventions reduced total activity in a graded way, with the thinking-enabled condition reducing activity to roughly two actions per active agent per active episode (\Cref{tab:engagement-ladder}; \Cref{fig:engagement-case-study}a).

Crucially, the interventions were not neutral corrections. As intervention strength increased, recommendation-algorithm differences narrowed or disappeared, and action composition shifted away from posting toward lighter interactions such as likes and replies (\Cref{fig:engagement-case-study}b). Finally, comparing Qwen3.5-4B and Qwen3.5-9B showed that model scale also moderated engagement dynamics: the larger model produced more activity in the relaxed-budget and strict-intervention settings, and preserved stronger recommendation-system effects under action pressure (\Cref{fig:engagement-case-study}c; \Cref{fig:action_dist_compare}).

Taken together, the engagement study shows why LLM social simulations require structured, component-wise experimentation. A monolithic analysis might conclude that recommendation systems do not affect engagement, that TwHIN increases engagement, or that realism prompts solve over-action. The study-driven workflow instead reveals a more precise conclusion: timeline curation can increase engagement, but the effect is masked by action-budget saturation, attenuated by agent-level realism interventions, and moderated by model scale.

\subsection{Case Study 3 — Echo Chambers: reproduction, elaboration, and extension of an existing study.}

A key target for moving toward a science of LLM-based simulations is enabling low-friction reproducibility \cite{anonymous2026position}. This case study illustrates how an existing LLM social simulation can be reproduced inside EASE, exposing which simulator assumptions are responsible for the reported phenomenon. We reconstruct the echo-chamber simulation of \citet{wang-etal-2025-decoding}, which studies polarization in LLM agent networks across scale-free, small-world, and random graph structures. In EASE, this simulator decomposes naturally: graph topology and recommendation policy belong to the environment, observation filtering to the game-master interface, short- and long-term belief processing to the agent, repeated social days to the simulation engine, and polarization, neighbor correlation, and disagreement to evaluation. The full study, including reproduction deltas and robustness tables, is provided in \Cref{app:echo_chambers}.
\vspace{-5pt}

\paragraph{Stage 1: Reproduction} We first attempt to reproduce the paper's core findings inside \textit{SiliSocS}, without changing the substantive simulator assumptions. The scale-free condition closely matches the original paper: it reports $\Delta$polarization $= +0.584$, $\Delta$NCI $= +0.670$, and $\Delta$global disagreement $= -0.471$, while our EASE reproduction obtains $+0.590$, $+0.579$, and $-0.362$, respectively (\Cref{tab:echo-reproduction}). Thus, the primary echo-chamber signature is reproduced: agents become more polarized, more locally correlated, and less disagreeing with their neighbors. Some graph-level magnitudes diverge, especially in the small-world and random settings, so we treat this as qualitative reproduction rather than exact trajectory-level replay. The core mechanism nevertheless survives translation into EASE.
\vspace{-5pt}

\paragraph{Stage 2: Follow-up} We then use the study schema to move beyond reproduction. Because the simulator has been decomposed, each follow-up can vary one component while holding the rest fixed. Changing observation formation from similarity exposure to opposing or random exposure reduces population-level polarization strongly, as expected, but does not straightforwardly reduce neighbor correlation \ref{app:opposing}. We then loosen the simulation engine and environment assumptions by replacing direct opinion exchange with a Twitter-like environment where agents can simultaneously post, repost, reply, and like before answering a structured belief probe. The echo-chamber pattern persists, with final polarization $2.990 \pm 0.150$, NCI $0.411 \pm 0.108$, and global disagreement $2.296 \pm 0.085$ (\Cref{tab:echo-loose-social}), indicating that the result can be robustly replicated even while generalizing the environment and engines.
\vspace{-5pt}

\paragraph{Stage 3: Robustness} Finally, we vary the agent and model assumptions made by the original work. Removing self-state feedback, ie, the agent's memory of its past decisions, lowers final polarization from $2.990$ to $2.695$ and increases belief volatility from $0.100$ to $0.238$, suggesting that these memories can act as anchoring mechanisms. Replacing the EchoSim short/long-memory agent with a simpler observe-memory-act agent further weakens echo-chamber behavior, lowering final polarization to $2.641$ and NCI to $0.193$ with self-state feedback, and to $2.391$ and $0.120$ without it (\Cref{tab:echo-loose-social}). Replacing GPT-4o-mini with Qwen3.5-4B changes the qualitative signature more strongly: Qwen agents show belief movement, but not the same positive local-neighborhood-correlation pattern, with Echo-style memory and self-state feedback yielding final polarization $2.605$ but NCI $-0.110$. This case study motivates the broader paper in two ways. First, EASE can serve as a reproduction layer for existing LLM simulations, making prior assumptions explicit rather than embedded in a monolithic codebase. Second, reproduction alone is insufficient. Once the simulator is expressed through configurable components, the study schema turns a single reported phenomenon into falsifiable design questions: is the effect carried by the graph, recommender, action interface, memory architecture, self-state prompt, or LLM backbone? The answer is distributed across these components. This is the central argument of EASE: modular simulation design enables bookkeeping over simulator assumptions, alongside lower-friction reproducibility.

\vspace{-5pt}
\section{Limitations}
\vspace{-5pt}
Our work provides an initial framework for modular simulation design and hypothesis-driven exploration, but it does not solve the broader validation problem for LLM-based social simulation. Many social phenomena lack ground-truth trajectories or agreed-upon behavioral metrics, and synthetic scenarios cannot by themselves establish real-world fidelity. EASE makes evaluation choices explicit, but the validity of any conclusion still depends on whether the selected metrics capture the phenomenon of interest.

A second limitation is that explicit configuration does not guarantee a valid configuration. The benefit of EASE is that assumptions are exposed and can be audited, including against methodological principles such as PIMMUR \citep{zhou2025pimmur}. However, users can still construct scenarios, prompts, or evaluations that leak the desired outcome, overfit to a metric, or otherwise induce the phenomenon under study. We therefore view configuration transparency as infrastructure for methodological scrutiny, not a replacement for it.

The design space of LLM social simulations also remains large. Although the study schema supports sequential hypothesis testing, exhaustive exploration across models, agents, environments, scenarios, seeds, and evaluation choices is often computationally infeasible. Our case studies illustrate how structured branching can make this process more tractable, but substantial computational resources may still be required.

Finally, the case studies should be interpreted as demonstrations of the EASE workflow rather than definitive empirical claims about social media, engagement, or polarization. Findings from LLM simulations must ultimately be compared against real-world data or independently validated experimental settings before being used to support strong claims about human social behavior. Recent works, such as \citet{anonymous2026position} have further discussed similar challenges.

\vspace{-5pt}
\section{Discussion}
\vspace{-5pt}
In this paper, we propose the EASE framework, decomposing the components of a social simulation to facilitate a science of social simulations. These include the agents, the environment, the simulation engine that orchestrates the interaction between the two, the scenario (the social/event context), and the evaluations. The configuration of these components is made straightforward in text-based configuration files (see \Cref{fig:schematic}). Scenarios and Evaluations are deployed in service of research questions that we structure into a hypothesis-centered study schema. The latter allows for sequential refinement and testing of follow-up hypotheses as suggested by the evaluation of the results. 

We provided three example case studies. In the first, we tackled the problem of style diversity, showing strong effects of the model strength, and revealing a subtlety in diversity: typical lexical diversity metrics hide diversity in stance. The latter, rather than the former, was influenced by having richer agent personas. In the second study, we explored how agents' engagement dynamics were affected by variation in the recommendation algorithm, user-action instructions, and model choice. We found that models showed more engagement when using a recommendation algorithm, with the effect being significant for the 9B model even under action constraint pressure. Additionally, we discover that models change their action distributions when pressured to become more realistic/reduce their actions. In the third study, we reproduce existing work by \cite{wang-etal-2025-decoding} on Echo-chamber formation and show that EASE and the study structure implemented via the \textit{SiliSocS} codebase can allow us to conduct ablations to gain insights in a structured manner. Together, these studies illustrate the kind of hypothesis-driven exploration of complex social simulations that EASE makes easy. Like all powerful technologies, LLMs have dual-use nature. Here, an accurate social simulator could be misused to design manipulation campaigns, persuasive advertising, engineered misinformation, or content exploiting social media algorithms.  However, 
these simulators would provide safe testing environments for developing defenses without harming anyone. Pursued responsibly, 
EASE allows the research community to ground and deepen our understanding of this fascinating new form of simulation of complex social systems.



\printbibliography

@misc{coppolillo2025engagementdrivencontentgenerationlarge,
      title={Engagement-Driven Content Generation with Large Language Models}, 
      author={Erica Coppolillo and Federico Cinus and Marco Minici and Francesco Bonchi and Giuseppe Manco},
      year={2025},
      eprint={2411.13187},
      archivePrefix={arXiv},
      primaryClass={cs.LG},
      url={https://arxiv.org/abs/2411.13187}, 
}

@misc{vallinder2024culturalevolutioncooperationllm,
      title={Cultural Evolution of Cooperation among LLM Agents}, 
      author={Aron Vallinder and Edward Hughes},
      year={2024},
      eprint={2412.10270},
      archivePrefix={arXiv},
      primaryClass={cs.MA},
      url={https://arxiv.org/abs/2412.10270}, 
}

@misc{shapira2026agentschaos,
      title={Agents of Chaos}, 
      author={Natalie Shapira and Chris Wendler and Avery Yen and Gabriele Sarti and Koyena Pal and Olivia Floody and Adam Belfki and Alex Loftus and Aditya Ratan Jannali and Nikhil Prakash and Jasmine Cui and Giordano Rogers and Jannik Brinkmann and Can Rager and Amir Zur and Michael Ripa and Aruna Sankaranarayanan and David Atkinson and Rohit Gandikota and Jaden Fiotto-Kaufman and EunJeong Hwang and Hadas Orgad and P Sam Sahil and Negev Taglicht and Tomer Shabtay and Atai Ambus and Nitay Alon and Shiri Oron and Ayelet Gordon-Tapiero and Yotam Kaplan and Vered Shwartz and Tamar Rott Shaham and Christoph Riedl and Reuth Mirsky and Maarten Sap and David Manheim and Tomer Ullman and David Bau},
      year={2026},
      eprint={2602.20021},
      archivePrefix={arXiv},
      primaryClass={cs.AI},
      url={https://arxiv.org/abs/2602.20021}, 
}

@inproceedings{El_Kishky_2022, series={KDD ’22},
   title={TwHIN: Embedding the Twitter Heterogeneous Information Network for Personalized Recommendation},
   url={http://dx.doi.org/10.1145/3534678.3539080},
   DOI={10.1145/3534678.3539080},
   booktitle={Proceedings of the 28th ACM SIGKDD Conference on Knowledge Discovery and Data Mining},
   publisher={ACM},
   author={El-Kishky, Ahmed and Markovich, Thomas and Park, Serim and Verma, Chetan and Kim, Baekjin and Eskander, Ramy and Malkov, Yury and Portman, Frank and Samaniego, Sofía and Xiao, Ying and Haghighi, Aria},
   year={2022},
   month=aug, pages={2842–2850},
   collection={KDD ’22} }

@misc{xi2023risepotentiallargelanguage,
      title={The Rise and Potential of Large Language Model Based Agents: A Survey}, 
      author={Zhiheng Xi and Wenxiang Chen and Xin Guo and Wei He and Yiwen Ding and Boyang Hong and Ming Zhang and Junzhe Wang and Senjie Jin and Enyu Zhou and Rui Zheng and Xiaoran Fan and Xiao Wang and Limao Xiong and Yuhao Zhou and Weiran Wang and Changhao Jiang and Yicheng Zou and Xiangyang Liu and Zhangyue Yin and Shihan Dou and Rongxiang Weng and Wensen Cheng and Qi Zhang and Wenjuan Qin and Yongyan Zheng and Xipeng Qiu and Xuanjing Huang and Tao Gui},
      year={2023},
      eprint={2309.07864},
      archivePrefix={arXiv},
      primaryClass={cs.AI},
      url={https://arxiv.org/abs/2309.07864}, 
}

@misc{bückkaeffer2026textitsiliconsocietycookbookdesign,
      title={The $\textit{Silicon Society}$ Cookbook: Design Space of LLM-based Social Simulations}, 
      author={Aurélien Bück-Kaeffer and Sneheel Sarangi and Maximilian Puelma Touzel and Reihaneh Rabbany and Zachary Yang and Jean-François Godbout},
      year={2026},
      eprint={2605.00197},
      archivePrefix={arXiv},
      primaryClass={cs.MA},
      url={https://arxiv.org/abs/2605.00197}, 
}

@inproceedings{
anonymous2026position,
title={Position: Time to Close The Validation Gap in {LLM} Social Simulations},
author={Maximilian {Puelma Touzel} and Sneheel Sarangi and Aurélien Bück-Kaeffer and Zachary Yang and Jean-François Godbout and Reihaneh Rabbany},
booktitle={Forty-third International Conference on Machine Learning Position Paper Track},
year={2026},
url={https://openreview.net/forum?id=LpbxLBcOBf}
}

@misc{qwen3.5,
    title  = {{Qwen3.5}: Towards Native Multimodal Agents},
    author = {{Qwen Team}},
    month  = {February},
    year   = {2026},
    url    = {https://qwen.ai/blog?id=qwen3.5}
}

@misc{mi2025mfllmsimulatingpopulationdecision,
      title={MF-LLM: Simulating Population Decision Dynamics via a Mean-Field Large Language Model Framework}, 
      author={Qirui Mi and Mengyue Yang and Xiangning Yu and Zhiyu Zhao and Cheng Deng and Bo An and Haifeng Zhang and Xu Chen and Jun Wang},
      year={2025},
      eprint={2504.21582},
      archivePrefix={arXiv},
      primaryClass={cs.MA},
      url={https://arxiv.org/abs/2504.21582}, 
}

@misc{venkit2026needsociallygroundedpersonaframework,
      title={The Need for a Socially-Grounded Persona Framework for User Simulation}, 
      author={Pranav Narayanan Venkit and Yu Li and Yada Pruksachatkun and Chien-Sheng Wu},
      year={2026},
      eprint={2601.07110},
      archivePrefix={arXiv},
      primaryClass={cs.CL},
      url={https://arxiv.org/abs/2601.07110}, 
}

@misc{ferrarotti2026generativeaicollectivebehavior,
      title={Generative AI collective behavior needs an interactionist paradigm}, 
      author={Laura Ferrarotti and Gian Maria Campedelli and Roberto Dessì and Andrea Baronchelli and Giovanni Iacca and Kathleen M. Carley and Alex Pentland and Joel Z. Leibo and James Evans and Bruno Lepri},
      year={2026},
      eprint={2601.10567},
      archivePrefix={arXiv},
      primaryClass={cs.AI},
      url={https://arxiv.org/abs/2601.10567}, 
}

@inproceedings{puelmatouzel2025,
  title     = {SandboxSocial: A Sandbox for Social Media Using Multimodal AI Agents},
  author    = {Puelma Touzel, Maximilian and Sarangi, Sneheel and Krishnakumar, Gayatri and Gurbuz, Busra Tugce and Welch, Austin and Yang, Zachary and Musulan, Andreea and Yu, Hao and Kosak-Hine, Ethan and Gibbs, Tom and Thibault, Camille and Rabbany, Reihaneh and Godbout, Jean-François and Zhao, Dan and Pelrine, Kellin},
  booktitle = {Proceedings of the Thirty-Fourth International Joint Conference on
               Artificial Intelligence, {IJCAI-25}},
  publisher = {International Joint Conferences on Artificial Intelligence Organization},
  editor    = {James Kwok},
  pages     = {11100--11103},
  year      = {2025},
  month     = {8},
  note      = {Demo Track},
  doi       = {10.24963/ijcai.2025/1271},
  url       = {https://doi.org/10.24963/ijcai.2025/1271},
}

@inproceedings{smith2024the,
title={The Concordia Contest: Advancing the Cooperative Intelligence of Language Agents},
author={Chandler Smith and Rakshit Trivedi and Jesse Clifton and Lewis Hammond and Akbir Khan and Sasha Vezhnevets and John P Agapiou and Edgar A. Du{\'e}{\~n}ez-Guzm{\'a}n and Jayd Matyas and Danny Karmon and Marwa Abdulhai and Dylan Hadfield-Menell and Natasha Jaques and Joel Z Leibo and Oliver Slumbers and Tim Baarslag and Minsuk Chang},
booktitle={NeurIPS 2024 Competition Track},
year={2024},
url={https://openreview.net/forum?id=dfeFy1PSSw}
}

@misc{barrie2025emergentllmbehaviorsobservationally,
      title={Emergent LLM behaviors are observationally equivalent to data leakage}, 
      author={Christopher Barrie and Petter Törnberg},
      year={2025},
      eprint={2505.23796},
      archivePrefix={arXiv},
      primaryClass={cs.CL},
      url={https://arxiv.org/abs/2505.23796}, 
}

@misc{zhou2025pimmur,
      title={The PIMMUR Principles: Ensuring Validity in Collective Behavior of LLM Societies}, 
      author={Jiaxu Zhou and Jen-tse Huang and Xuhui Zhou and Man Ho Lam and Xintao Wang and Hao Zhu and Wenxuan Wang and Maarten Sap},
      year={2025},
      eprint={2509.18052},
      archivePrefix={arXiv},
      primaryClass={cs.CL},
      url={https://arxiv.org/abs/2509.18052}, 
}

@misc{piao2025agentsocietylargescalesimulationllmdriven,
      title={AgentSociety: Large-Scale Simulation of LLM-Driven Generative Agents Advances Understanding of Human Behaviors and Society}, 
      author={Jinghua Piao and Yuwei Yan and Jun Zhang and Nian Li and Junbo Yan and Xiaochong Lan and Zhihong Lu and Zhiheng Zheng and Jing Yi Wang and Di Zhou and Chen Gao and Fengli Xu and Fang Zhang and Ke Rong and Jun Su and Yong Li},
      year={2025},
      eprint={2502.08691},
      archivePrefix={arXiv},
      primaryClass={cs.SI},
      url={https://arxiv.org/abs/2502.08691}, 
}

@misc{larooij2025largelanguagemodelssolve,
      title={Do Large Language Models Solve the Problems of Agent-Based Modeling? A Critical Review of Generative Social Simulations}, 
      author={Maik Larooij and Petter Törnberg},
      year={2025},
      eprint={2504.03274},
      archivePrefix={arXiv},
      primaryClass={cs.MA},
      url={https://arxiv.org/abs/2504.03274}, 
}

@inproceedings{ng2025llm,
  title={Are LLM-Powered Social Media Bots Realistic?},
  author={Ng, Lynnette Hui Xian and Carley, Kathleen M},
  booktitle={International Conference on Social Computing, Behavioral-Cultural Modeling and Prediction and Behavior Representation in Modeling and Simulation},
  pages={14--23},
  year={2025},
  organization={Springer}
}

@inproceedings{
anthis2025position,
title={Position: {LLM} Social Simulations Are a Promising Research Method},
author={Jacy Reese Anthis and Ryan Liu and Sean M Richardson and Austin C. Kozlowski and Bernard Koch and Erik Brynjolfsson and James Evans and Michael S. Bernstein},
booktitle={Forty-second International Conference on Machine Learning Position Paper Track},
year={2025},
url={https://openreview.net/forum?id=cRBg1dtj7o}
}

@misc{zhou2025socialworldmodels,
      title={Social World Models}, 
      author={Xuhui Zhou and Jiarui Liu and Akhila Yerukola and Hyunwoo Kim and Maarten Sap},
      year={2025},
      eprint={2509.00559},
      archivePrefix={arXiv},
      primaryClass={cs.AI},
      url={https://arxiv.org/abs/2509.00559}, 
}

@misc{vezhnevets2025multiactor,
      title={Multi-Actor Generative Artificial Intelligence as a Game Engine}, 
      author={Alexander Sasha Vezhnevets and Jayd Matyas and Logan Cross and Davide Paglieri and Minsuk Chang and William A. Cunningham and Simon Osindero and William S. Isaac and Joel Z. Leibo},
      year={2025},
      eprint={2507.08892},
      archivePrefix={arXiv},
      primaryClass={cs.AI},
      url={https://arxiv.org/abs/2507.08892}, 

}

@inproceedings{buckkaeffer2025blueprint,
  title={BluePrint: A Social Media User Dataset for LLM Persona Evaluation and Training},
  author={B{\"u}ck-Kaeffer, Aur{\'e}lien and Chooi, Je Qin and Zhao, Dan and Puelma Touzel, Maximilian and Pelrine, Kellin and Godbout, Jean-Fran{\c{c}}ois and Rabbany, Reihaneh and Yang, Zachary},
  booktitle={Workshop on Tailoring AI: Exploring Active and Passive LLM Personalization (PALS)},
  year={2025},
  organization={EMNLP},
url={https://pals-nlp-workshop.github.io/}
}

@inproceedings{wang-etal-2025-decoding,
    title = "Decoding Echo Chambers: {LLM}-Powered Simulations Revealing Polarization in Social Networks",
    author = "Wang, Chenxi  and
      Liu, Zongfang  and
      Yang, Dequan  and
      Chen, Xiuying",
    editor = "Rambow, Owen  and
      Wanner, Leo  and
      Apidianaki, Marianna  and
      Al-Khalifa, Hend  and
      Eugenio, Barbara Di  and
      Schockaert, Steven",
    booktitle = "Proceedings of the 31st International Conference on Computational Linguistics",
    month = jan,
    year = "2025",
    address = "Abu Dhabi, UAE",
    publisher = "Association for Computational Linguistics",
    url = "https://aclanthology.org/2025.coling-main.264/",
    pages = "3913--3923"
}

@misc{zuo2025mtosllmdrivenmultitopicopinion,
      title={MTOS: A LLM-Driven Multi-topic Opinion Simulation Framework for Exploring Echo Chamber Dynamics}, 
      author={Dingyi Zuo and Hongjie Zhang and Jie Ou and Chaosheng Feng and Shuwan Liu},
      year={2025},
      eprint={2510.12423},
      archivePrefix={arXiv},
      primaryClass={cs.AI},
      url={https://arxiv.org/abs/2510.12423}, 
}

@inproceedings{Piatti2024,
 author = {Piatti, Giorgio and Jin, Zhijing and Kleiman-Weiner, Max and Sch\"{o}lkopf, Bernhard and Sachan, Mrinmaya and Mihalcea, Rada},
 booktitle = {Advances in Neural Information Processing Systems},
 editor = {A. Globerson and L. Mackey and D. Belgrave and A. Fan and U. Paquet and J. Tomczak and C. Zhang},
 pages = {111715--111759},
 publisher = {Curran Associates, Inc.},
 title = {Cooperate or Collapse:  Emergence of Sustainable Cooperation in a Society of LLM Agents},
 url = {https://proceedings.neurips.cc/paper_files/paper/2024/file/ca9567d8ef6b2ea2da0d7eed57b933ee-Paper-Conference.pdf},
 volume = {37},
 year = {2024}
}

@misc{park2022socialsimulacracreatingpopulated,
      title={Social Simulacra: Creating Populated Prototypes for Social Computing Systems}, 
      author={Joon Sung Park and Lindsay Popowski and Carrie J. Cai and Meredith Ringel Morris and Percy Liang and Michael S. Bernstein},
      year={2022},
      eprint={2208.04024},
      archivePrefix={arXiv},
      primaryClass={cs.HC},
      url={https://arxiv.org/abs/2208.04024}, 
}

@misc{yalabadi2024controllingmisinformationdiffusionsocial,
      title={Controlling the Misinformation Diffusion in Social Media by the Effect of Different Classes of Agents}, 
      author={Ali Khodabandeh Yalabadi and Mehdi Yazdani-Jahromi and Sina Abdidizaji and Ivan Garibay and Ozlem Ozmen Garibay},
      year={2024},
      eprint={2401.11524},
      archivePrefix={arXiv},
      primaryClass={cs.MA},
      url={https://arxiv.org/abs/2401.11524}, 
}

@misc{orlando2025emergentcoordinatedbehaviorsnetworked,
      title={Emergent Coordinated Behaviors in Networked LLM Agents: Modeling the Strategic Dynamics of Information Operations}, 
      author={Gian Marco Orlando and Jinyi Ye and Valerio La Gatta and Mahdi Saeedi and Vincenzo Moscato and Emilio Ferrara and Luca Luceri},
      year={2025},
      eprint={2510.25003},
      archivePrefix={arXiv},
      primaryClass={cs.MA},
      url={https://arxiv.org/abs/2510.25003}, 
}

@misc{ding2025understandingworldpredictingfuture,
      title={Understanding World or Predicting Future? A Comprehensive Survey of World Models}, 
      author={Jingtao Ding and Yunke Zhang and Yu Shang and Yuheng Zhang and Zefang Zong and Jie Feng and Yuan Yuan and Hongyuan Su and Nian Li and Nicholas Sukiennik and Fengli Xu and Yong Li},
      year={2025},
      eprint={2411.14499},
      archivePrefix={arXiv},
      primaryClass={cs.CL},
      url={https://arxiv.org/abs/2411.14499}, 
}

@misc{larooij2025fixsocialmediatesting,
      title={Can We Fix Social Media? Testing Prosocial Interventions using Generative Social Simulation}, 
      author={Maik Larooij and Petter Törnberg},
      year={2025},
      eprint={2508.03385},
      archivePrefix={arXiv},
      primaryClass={cs.SI},
      url={https://arxiv.org/abs/2508.03385}, 
}

@misc{neumann2025usellmssimulateopinions,
      title={Should you use LLMs to simulate opinions? Quality checks for early-stage deliberation}, 
      author={Terrence Neumann and Maria De-Arteaga and Sina Fazelpour},
      year={2025},
      eprint={2504.08954},
      archivePrefix={arXiv},
      primaryClass={cs.CY},
      url={https://arxiv.org/abs/2504.08954}, 
}

@article{yang2024oasis,
  title={Oasis: Open agent social interaction simulations with one million agents},
  author={Yang, Ziyi and Zhang, Zaibin and Zheng, Zirui and Jiang, Yuxian and Gan, Ziyue and Wang, Zhiyu and Ling, Zijian and Chen, Jinsong and Ma, Martz and Dong, Bowen and others},
  journal={arXiv preprint arXiv:2411.11581},
  year={2024}
}

@misc{vezhnevets2023generativeagentbasedmodelingactions,
      title={Generative agent-based modeling with actions grounded in physical, social, or digital space using Concordia}, 
      author={Alexander Sasha Vezhnevets and John P. Agapiou and Avia Aharon and Ron Ziv and Jayd Matyas and Edgar A. Duéñez-Guzmán and William A. Cunningham and Simon Osindero and Danny Karmon and Joel Z. Leibo},
      year={2023},
      eprint={2312.03664},
      archivePrefix={arXiv},
      primaryClass={cs.AI},
      url={https://arxiv.org/abs/2312.03664}, 
}

@misc{zhou2026pimmur,
      title={The PIMMUR Principles: Ensuring Validity in Collective Behavior of LLM Societies}, 
      author={Jiaxu Zhou and Jen-tse Huang and Xuhui Zhou and Man Ho Lam and Xintao Wang and Hao Zhu and Wenxuan Wang and Maarten Sap},
      year={2026},
      eprint={2509.18052},
      archivePrefix={arXiv},
      primaryClass={cs.CL},
      url={https://arxiv.org/abs/2509.18052}, 
}

@misc{leibo2026theoryappropriatenessaccountsnorms,
      title={A Theory of Appropriateness That Accounts for Norms of Rationality}, 
      author={Joel Z. Leibo and Alexander Sasha Vezhnevets and Manfred Diaz and John P. Agapiou and William A. Cunningham and Peter Sunehag and Logan Cross and Raphael Koster and Stanley M. Bileschi and Minsuk Chang and Iyad Rahwan and Simon Osindero and James A. Evans},
      year={2026},
      eprint={2603.14050},
      archivePrefix={arXiv},
      primaryClass={cs.NE},
      url={https://arxiv.org/abs/2603.14050}, 
}

@software{nemotron,
  author = {Meyer, Yev and Corneil, Dane},
  title = {{Nemotron-Personas-USA}: Synthetic Personas Aligned to Real-World Distributions
},
  month = {June},
  year = {2025},
  url = {https://huggingface.co/datasets/nvidia/Nemotron-Personas-USA}
}

@inproceedings{design,
author = {Sangiovanni-Vincentelli, Alberto and Carloni, Luca and De Bernardinis, Fernando and Sgroi, Marco},
title = {Benefits and challenges for platform-based design},
year = {2004},
isbn = {1581138288},
publisher = {Association for Computing Machinery},
address = {New York, NY, USA},
url = {https://doi.org/10.1145/996566.996684},
doi = {10.1145/996566.996684},
booktitle = {Proceedings of the 41st Annual Design Automation Conference},
pages = {409–414},
numpages = {6},
location = {San Diego, CA, USA},
series = {DAC '04}
}


\appendix

\paragraph{LLM Disclosure Statement}

In this paper, we used the GPT-5.3-Codex model via GitHub Copilot to interpret data and help in the generation of some case study plots.

\section{Reference Scenarios}
\label{app:refscen}
\paragraph{Fake Election}

This scenario is used in our engagement dynamics case study. It follows a mayoral election in a small town named Storhampton, with two candidates, Bill and Bradley, who have policies that are slightly conservative and slightly progressive leaning respectively. 

Beyond general scenario information, we create the profiles of the candidates and 3 News-streams: biased towards Bill, biased towards Bradley, and Neutral. In this scenario, we encourage users to use pre-existing persona datasets such as the Nemotron-Personas-USA \citep{nemotron} dataset for the voter agents.

\paragraph{AI Conference}
This scenario is used in the style diversity study. It consists of two groups (3 protesters, 2 journalists, and 4 conference attendees) outside and inside the same event: an AI industry conference. 

\paragraph{Misinformation}
This scenario is used in the style diversity study. A small social media community of six users with distinct personalities and habits, creating an asymmetric information flow. A false health claim has been posted on the platform claiming that a common household item causes serious health issues.

\section{Codebase Details}
\label{app:codebase}

We provide default components that should cover the design of a large number of simulation structures purely through the configuration interface. 

\textbf{Environment:} we provide 3 social media environment objects: twitter-like, reddit-like, and a mastodon object that can be used to have agents interact with a hosted mastodon serve. Additionally, we provide default game-masters which function with this environment. 

\textbf{Agent:} We share language model objects for models served in the OpenAI format, a basic agent architecture, a persona pipeline to pull personas from publicly available datasets as well as locally present .json files, and a simple list-based memory system. We also allow for configurability such that users can use different LoRA models to serve different agents. 

\textbf{Simulation Engine:} We share a simultaneous engine object that lets agents act parallelly. We provide three agent action policy settings: Single Agent action/episode, fixed number of agent actions per episode, and an agent-decided number of actions per episode with a maximum cap. 

\textbf{Evaluations:} We provide easily configurable probe objects that can be provided to LLM agents during the simulation. Additionally, we provide a basic dashboard for post-simulation analysis.

In general, as our framework builds upon the Concordia \citep{vezhnevets2023generativeagentbasedmodelingactions} framework, users can borrow components and designs foro agents, game-masters, engines etc. from the Concordia library, and use them in our framework with minimal modifications.

\paragraph{Extending to Non-Social Media Environments} \label{app:nonsm} Although our case studies use a social-media-like backend, the backend is not fixed. The environment is implemented as a replaceable object that exposes methods and argument signatures to the game master and agents. The game master uses this interface to form observations, validate intended actions, and resolve actions against the backend state. Any other interaction environment can therefore be freely substituted while preserving the same agent, engine, and evaluation abstractions. However, in practice, it may be necessary to slightly modify other components of a simulation to better support performance in a given backend environment.

\section{Case Study Details}
\subsection{Style Diversity}\label{sec:diversity_details}
The study built off of the misinformation and AI conference scenarios.
\subsubsection{Diversity Metrics}\label{sec:div_metrics}
We focus on three metrics that each capture a distinct dimension of linguistic diversity. All are computed per-agent from their post history (excluding boosts), then averaged.







\paragraph{Lexical Diversity (higher = more diverse)}

The classic type--token ratio (TTR): the number of unique word types divided by the total number of word tokens across all of an agent's posts. Tokenization uses a simple \verb|\w+| regex on lowercased text.

\emph{Exact definition:}

\begin{equation*}
\text{Lexical Diversity}_a = \frac{|\{\text{unique tokens}\}|}{|\text{all tokens}|}
\end{equation*}

over all posts by agent $a$. Ranges from $1/N$ (one word repeated $N$ times) to $1.0$ (every word unique).

\textbf{Why it matters:} TTR captures vocabulary richness. A low score means the agent relies on a small working vocabulary, a hallmark of repetitive LLM output. It complements self-BLEU by measuring diversity at the single-word level rather than the phrase level.

\paragraph{Opener Variety (higher = more diverse)}

The fraction of an agent's posts that begin with a unique 5-word prefix. Each post's first 5 tokens are extracted as a tuple; opener variety = (number of unique openers) / (number of posts).

\emph{Exact definition:}

\begin{equation*}
\text{Opener Variety}_a = \frac{|\{\text{unique 5-word prefixes}\}|}{|\text{posts}|}
\end{equation*}

A score of 1.0 means every post starts differently; a score of $1/N$ means all $N$ posts open identically.

\textbf{Why it matters:} Formulaic openings (``That's a great point, \ldots'') are a common LLM repetition pattern and are immediately noticeable to readers. This metric specifically targets that failure mode.

\paragraph{Inter-Agent Distinctiveness (higher = more diverse)}
For each agent, all of their posts are concatenated and a single term-frequency vector is computed. Inter-agent distinctiveness is the mean pairwise cosine distance across all pairs of agents.
\emph{Exact definition:}
\begin{equation*}
\text{Inter-Agent Distinctiveness} = \frac{1}{\binom{A}{2}} \sum_{i < j} \left(1 - \frac{\mathbf{tf}_i \cdot \mathbf{tf}_j}{\|\mathbf{tf}_i\|\;\|\mathbf{tf}_j\|}\right)
\end{equation*}
where $\mathbf{tf}_i$ is the aggregate term-frequency vector of all posts by agent $i$ and $A$ is the number of agents with at least 2 posts.

\textbf{Why it matters:} The per-agent metrics above measure self-diversity (does each agent repeat itself?). Inter-agent distinctiveness measures \emph{cross-agent} diversity---do different agents actually sound different from each other? A score near 0 means all agents converge on the same vocabulary (groupthink). A high score means each agent occupies a distinct linguistic niche, consistent with genuine persona differentiation.
\subsubsection{Survey responses}
\begin{figure}
    \centering
    \includegraphics[width=0.3\linewidth]{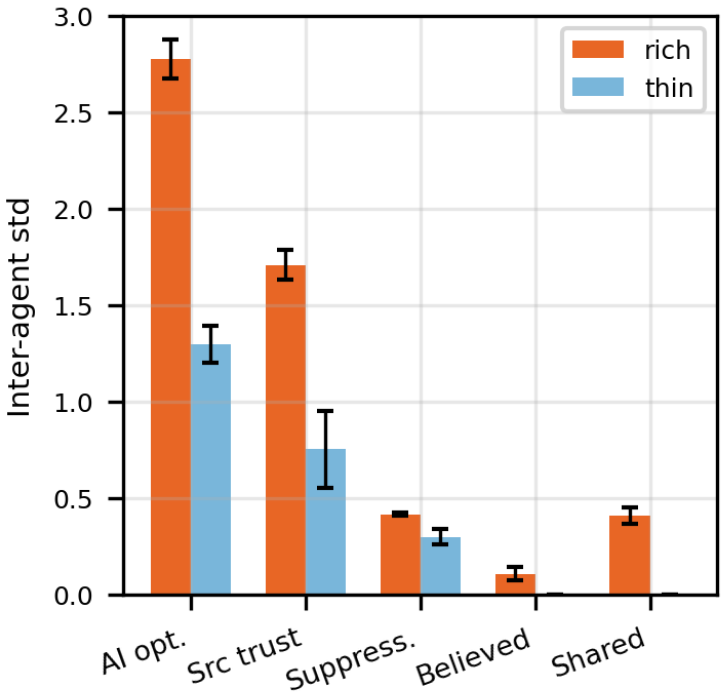}
    \caption{Score Diversity of responses to probe questions.}
    \label{fig:stance}
\end{figure}

We confirmed that the richer personas came through in expressing diversity across survey questions (\Cref{fig:stance}).
Beyond text diversity, agents are probed each step with structured questions about their current beliefs.
These measure whether agents hold \textbf{meaningfully different positions}---not just different words.

\textbf{Probe questions:}
\begin{itemize}
    \item \texttt{ai\_optimism} \emph{(1--10)}: ``How optimistic do you feel about AI's overall impact on society right now?''
    \item \texttt{source\_trust} \emph{(1--10)}: ``How much do you trust posts from accounts you don't personally know?''
    \item \texttt{suppressed\_concerns} \emph{(yes/no)}: ``Have you held back an opinion or concern because you worried how others would react?''
    \item \texttt{believed\_claim} \emph{(yes/no, misinformation)}: ``Do you currently believe that a common household item causes serious health issues?''
    \item \texttt{shared\_misinfo} \emph{(yes/no, misinformation)}: ``Have you posted or boosted content claiming a household item causes health issues?''
\end{itemize}

\textbf{Metric:} inter-agent std of each agent's mean response---higher = agents disagree more with each other.

\subsubsection{NLI Stance Divergence in Posts}

Zero-shot NLI classification of post \emph{content}---do agents take different stances in what they actually write?

Each post is scored against the scenario claim using \texttt{cross-encoder/nli-deberta-v3-small}:
\begin{itemize}
    \item \textbf{ai\_conference}: ``AI is beneficial and society should embrace it.''
    \item \textbf{misinformation}: ``A common household item causes serious health issues.''
\end{itemize}

Score = P(pro) $-$ P(anti), ranging from $-1$ (strongly anti) to $+1$ (strongly pro). \\

\subsection{Engagement Dynamics}
\label{app:engagement_dynamics}
This case study examines how timeline curation, action-budget constraints, model scale, and prompt/structure interventions jointly shape engagement dynamics.
We run each setting for n=10 seeds, and report seed-matched statistics as primary evidence wherever available, with Holm-adjusted p-values for within-family multiple comparisons.
The scenario fixes agent personas, background context, and timeline size (10 posts), and iterates over other simulation components. We choose this design primarily because, to the best of our knowledge, no previous work has explicitly studied the effects of allowing models to take multiple actions within a step, that simulated the burstiness of human action taking in social-media. Previous LLM Social simulation work such has often relied on allowing agents single actions, or a fixed number of actions within timesteps \citep{yang2024oasis, puelmatouzel2025}. We primarily use the Qwen3.5-4B \citep{qwen3.5} model in non-thinking mode as our base model.

\paragraph{H1. Timeline Curation Algorithms should lead to higher engagement}
Recommendation system algorithms are designed to elicit stronger human reactions, and keep users engaged with platforms. As such, it is plausible that agents in the simulation will choose to take more actions when served a curated algorithm-generated timeline. We set a maximum action cap of 12 actions per timestep for each agent and run the simulation.

\textbf{What was varied.} Timeline algorithms: 
\begin{enumerate}
    \item Follower-Chronological: retrieves the 10 most recent posts, replies, or reposts from followed users. 
    \item General embedding: Uses a general sentence-transformers model to retrieve the top 10 similar posts to the user's profile, which is generated by combining their persona description, 10 most recent posts, and 10 most recent liked posts.
    \item TwHIN Encoder: Same as above, but uses the TwHIN \citep{El_Kishky_2022} model that is trained on Twitter data to compute similarity. We borrow implementation details of the two recsys algorithms from OASIS \citep{yang2024oasis}.
\end{enumerate}

\textbf{Outcome.} We observe that total actions show no significant differences across the different timeline curation settings.

\textbf{Interpretation.} Even though we see no meaningful variation, we note that the mean actions across arms (\textgreater 11) appear close to the set cap of 12 actions. This weak finding motivates H2: 

\paragraph{H2. Increasing Action-Cap per-step will show discernible differences}
We reason that tight action bounds may mask the effects of timeline curation, as all agents hit the maximum action wall. Loosening this constraint, and in the process allowing the agent to take more actions, should give us better insights into H1.

\textbf{What was varied.} We keep the same variation as H1 (timeline curation) but increase the action budget per agent, per episode from 12, to 20.

\textbf{Outcome.} Relaxing budget increases all arm totals by $\sim 4$ actions per agent, and critical arm differences now emerge: recsys\_TwHIN (16.91) significantly exceeds chronological (15.60) on total actions ($p^{**}$) and interactions ($p^{**}$).

\textbf{Interpretation.} As hypothesized, budget constraint in H1 masked algorithmic separation; relaxing it reveals a small, but significant advantage for the TwHIN algorithm on total activity. 

However, given the model is only exposed to a fixed 10 post timeline, taking \textgreater 16 actions is unrealistic. We next hypothesize that it's possible to restrict the action count in LLMs by specific agent-level interventions.

\begin{table}[t]
  \centering
  \small
  \begin{tabular}{lrrr}
  \toprule
  \textbf{Regime} & \textbf{Chronological} & \textbf{recsys\_general} & \textbf{recsys\_TwHIN} \\
  \midrule
  \multicolumn{4}{c}{\textbf{Qwen3.5-4B}} \\
  \midrule
  Baseline & $11.21\pm0.40$ & $11.37\pm0.35$ & $\mathbf{11.41\pm0.56}$ \\
  Relaxed Budget & $15.60\pm0.83$ & $16.38\pm0.81$ & $\mathbf{16.91\pm1.01}^{**}$ \\
  Mild Intervention & $16.13\pm0.97$ & $\mathbf{17.35\pm0.63}^{*}$ & $16.81\pm1.49$ \\
  Moderate Intervention & $12.62\pm1.23$ & $13.17\pm0.80$ & $\mathbf{13.86\pm1.75}$ \\
  Strict Intervention & $10.08\pm0.80$ & $10.63\pm0.40$ & $\mathbf{10.70\pm0.88}$ \\
  Thinking-Enabled & $2.11\pm0.11$ & $2.09\pm0.17$ & $\mathbf{2.14\pm0.09}$ \\
  \midrule
  \multicolumn{4}{c}{\textbf{Qwen3.5-9B}} \\
  \midrule
  Relaxed Budget & $20.38\pm1.09$ & $\mathbf{21.15\pm0.91}$ & $20.47\pm3.68$ \\
  Strict & $12.95\pm1.25$ & $14.87\pm1.23^{**}$ & $\mathbf{16.24\pm1.47}^{**}$ \\
  Thinking-Enabled & $\mathbf{2.30\pm0.14}$ & $2.23\pm0.11$ & $2.27\pm0.12$ \\
  \bottomrule
  \end{tabular}
  \caption{Mean total actions per active-agent per active-episode (mean$\pm$sd) across regimes and timeline algorithms. Bold marks the largest mean within each row. Stars mark algorithms significantly different from chronological within the same regime (paired Wilcoxon on seeds with Holm correction: $^{*}p<0.05$, $^{**}p<0.01$).}
  \label{tab:engagement-ladder}
\end{table}

\paragraph{H3. Agent-level interventions can suppress action frequency} To gain more human-like action frequencies, we attempt to control LLM behavior via a range of interventions.

\textbf{What was varied.} For the Mild, Moderate, and Strict interventions, we append a prompt asking the model to consider the actions it takes and act realistically with increasing strictness. In the thinking setting, on top of the strict intervention prompt, we ask the model to reason about the actions it will take before we prompt it to take the action. We provide the detailed intervention prompts in Appendix \ref{app:intervention_prompts}.

\textbf{Outcome: Activity suppression.} In table~\ref{tab:engagement-ladder}, rows 3--6, we can observe that each intervention step, excluding the Mild intervention, reduces mean total actions: Moderate cuts $\sim 3$ actions, Strict cuts $\sim 2$ more, Thinking-Enabled significantly cuts down mean actions to $\sim 2$. Interestingly, the Mild prompt does not lead to a significant change from the Relaxed-Baseline.

\textbf{Outcome: Algorithmic erasure.} As intervention strictness increases, the recommendation-algorithm-based differences vanish. In H2, we showed TwHIN outperforming chronological by $\sim 1.3$ actions (significant). Under Moderate Intervention, this gap narrows to $\sim 1.2$, and is not significant in our setup due to the high variance. Under the Strict and Thinking-Enabled set-ups, the mean difference is reduced even further.

\begin{figure}[htbp]
    \centering
    \includegraphics[width=\textwidth]{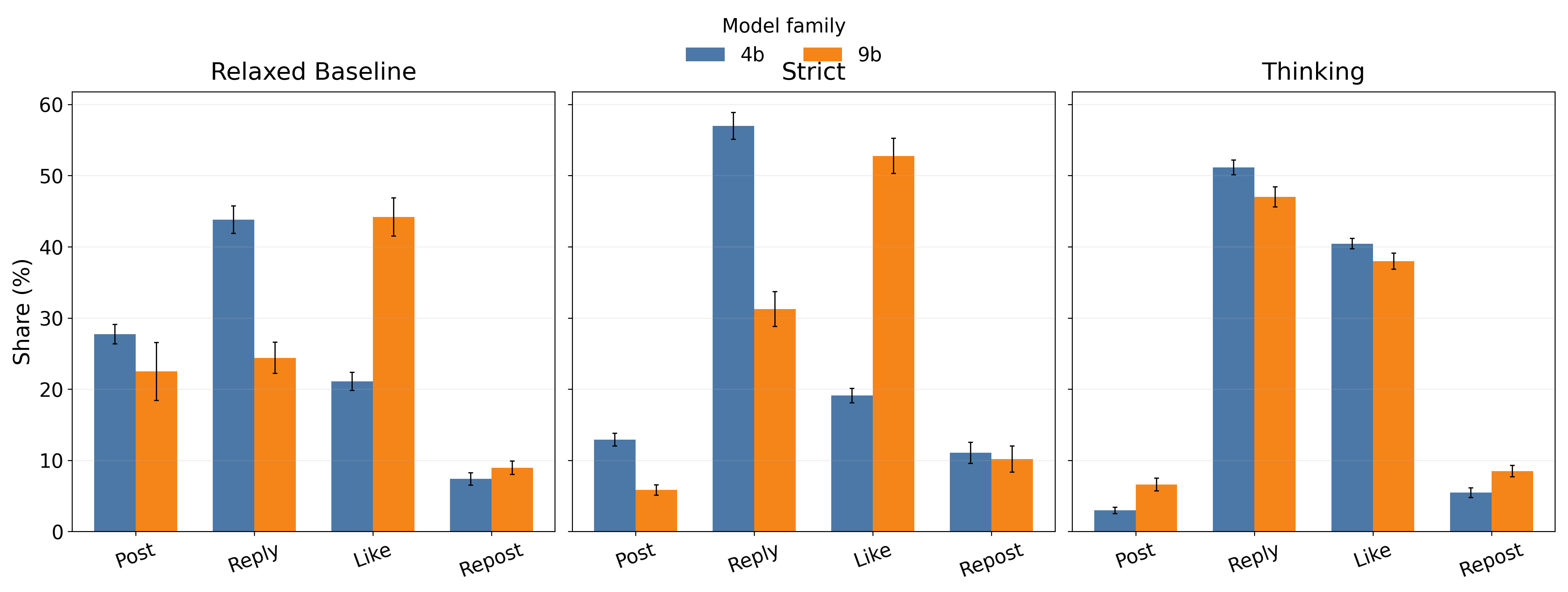}
    \caption{Action distribution differences between the Qwen3.5-4B and 9B model.}
    \label{fig:action_dist_compare}
\end{figure}

\textbf{Outcome: Action channel redistribution.} As shown in Figure \ref{fig:action_dist_compare}, interventions suppress action channels non-uniformly: post share declines monotonically from $27.71\%$ in the Relaxed Baseline setting to $2.97\%$ Thinking setting, while reply share rises from $43.81\%$ to $48.23\%$ and peaks at $56.98\%$ under the Strict setting. Like share is comparatively stable through the Moderate and Strict settings($\sim19$--$21\%$) but jumps to $40.44\%$ in the Thinking setting, indicating that deliberative constraints suppress posting/reply initiation much more than lightweight engagement actions. Since the prompts nudge the model to act realistically, this gives us insights into what LLMs consider realistic behavior. 


\textbf{Interpretation.} Prompt-based interventions are effective behavioral control levers: they can suppress activity in a graded, reliable manner, but need evaluations to truly judge effects. Additionally, prompting models to act in a more constrained, realistic manner does not preserve any recommendation algorithm-based effects. The strongest intervention (Thinking-Enabled) achieves near-perfect equivalence. Mechanistically, interventions compress the post/interaction ratio steeply, suggesting constraints focus on initiation rather than participation. This motivates H4: given that interventions so thoroughly suppress baseline effects, do \emph{fundamental} differences (model scale) still persist?

\paragraph{H4. Models at different scales diverge in engagement behavior.} Larger models are expected to be more capable. While it is unclear if general capability leads to better simulation capability, with some works suggesting the contrary \citep{mi2025mfllmsimulatingpopulationdecision}, we expect to see divergence in model behavior.

\textbf{What was varied.} LLM Backbone: between Qwen3.5-4B, and Qwen3.5-9B.

\textbf{Outcome: Divergence in Actions} In table~\ref{tab:engagement-ladder}, we observe that the 9B model consistently and significantly outperforms the 4B model on total actions across timeline algorithms and the Relaxed Budget, and Strict settings. The 9B model's advantage is concentrated in interaction (likes + reposts + replies) channels: the Relaxed Budget setting shows +4.27 interactions ($p^{**}$) but only +0.52 posts (ns); Strict Intervention shows +3.02 interactions ($p^{**}$) and $-0.15$ posts (ns). This narrow distribution indicates model scale amplifies response behaviors (replies, likes, reposts) rather than initiating new posts. 

Additionally, as shown in Figure \ref{fig:action_dist_compare}, the 9B model's action distribution varies significantly from the 4B model in that it produces a higher proportion of like-behavior and a lower proportion of reply behavior in both the Relaxed baseline and Strict settings, but matches the 4B algorithm more closely in the Thinking setting.

\textbf{Outcome: Algorithm Effects} We note that due to the saturation of the action budget, the Relaxed Budget setting for Qwen3.5-9B likely faces the same issues of hidden algorithmic effects that were highlighted in H1 for Qwen3.5-4B. Interestingly, however, we see a significantly higher total action count even in the Strict setting, where Qwen3.5-4B model observes a flattening of the algorithmic effects.

\textbf{Interpretation.} Model-scale differences can be observed across settings and algorithms. The 9B model consistently performs more actions than the 4B model, and has a different action distribution. Furthermore, the model seems to still engage more with a recommendation system algorithm compared to the baseline in a constrained action setting. This may indicate that the 9B model is better at modeling a persona's interests, under action constraint pressure.

\paragraph{Discussion}

The case study reveals several interesting behaviors of LLM agent models. Our findings indicate that LLM agents naturally tend to interact more with curated algorithms, thus suggesting some level of human-interest modeling capability, although the magnitude of the effect remains small. Additionally, we demonstrate that models' action behavior can effectively be controlled by prompt-based approaches, and in doing this, the model changes its action profile, shifting away from taking posts. Thus indicating that models have some concept of excessive posting being less realistic. Additionally, we observe that models of different scales, even when hailing from the same model family, can diverge significantly in behavior. In general, this sort of hypothesis-driven study becomes straightforward to conduct when a simulation system can be decomposed into its modular components. In Appendix \ref{app:hyp}, we present further interesting hypotheses and routes for this study.

\subsection{Case Study: Echo Chambers and Opinion Polarization}
\label{app:echo_chambers}

This case study attempts to reproduce and extend the LLM-based echo-chamber simulation of \citet{wang-etal-2025-decoding} using the EASE framework and the \texttt{SiliSocS} codebase. The original study shows that LLM agent networks can reproduce echo-chamber formation across social graph structures, especially scale-free and small-world graphs, and that similarity-based exposure can increase polarization and local neighborhood agreement. We use this case study to show how an existing simulator can first be re-expressed in a modular framework, and then systematically relaxed along specific simulator assumptions.

Unless otherwise noted, reproduction and follow-up experiments use GPT-4o-mini and are run for $n=5$ seeds. We use the paper-style metrics: polarization as belief variance, global disagreement as degree-normalized neighbor belief distance, and neighbor correlation index (NCI) as the correlation between each agent's belief and its neighborhood belief. Stronger echo-chamber formation is indicated by higher polarization, higher NCI, and lower global disagreement.

\paragraph{H1. EchoSim can be reproduced inside the EASE framework.}
We first reconstruct the original EchoSim design using EASE without changing the substantive simulator assumptions. The agent keeps the original persona, short-term memory, long-term memory, and belief-update structure. The environment owns the graph, belief state, and recommendation rule. The simulation engine runs repeated social days, and the evaluation layer computes polarization, NCI, and global disagreement.

\textbf{What was varied.} We reproduce the original graph comparison across small-world, scale-free, and random graphs under similarity-based exposure.

\textbf{Outcome.} The in-framework reproduction preserves the main qualitative result. The strongest match is the scale-free graph, where the original paper reports $\Delta$polarization $=+0.584$, $\Delta$NCI $=+0.670$, and $\Delta$global disagreement $=-0.471$. Our EASE reproduction obtains $\Delta$polarization $=+0.590$, $\Delta$NCI $=+0.579$, and $\Delta$global disagreement $=-0.362$. Thus, the primary echo-chamber signature is reproduced: agents become more polarized, more locally correlated, and less disagreeing with their neighbors.

\begin{table}[t]
\centering
\small
\begin{tabular}{lrrr}
\toprule
\textbf{Condition} & $\Delta$Polarization $\uparrow$ & $\Delta$Global Disagreement $\downarrow$ & $\Delta$NCI $\uparrow$ \\
\midrule
Paper, small-world & $+0.584$ & $-0.185$ & $+0.400$ \\
EASE, small-world & $+0.520$ & $+0.023$ & $+0.296$ \\
\midrule
Paper, scale-free & $+0.584$ & $-0.471$ & $+0.670$ \\
EASE, scale-free & $+0.590$ & $-0.362$ & $+0.579$ \\
\midrule
Paper, random & $+0.936$ & $+0.164$ & $+0.486$ \\
EASE, random & $+0.460$ & $+0.048$ & $+0.353$ \\
\bottomrule
\end{tabular}
\caption{Paper-reported deltas from \citet{wang-etal-2025-decoding} compared with our in-framework \textit{SiliSocS} reproduction.}
\label{tab:echo-reproduction}
\end{table}

\textbf{Divergence from the paper.} The reproduction is not numerically identical. In particular, the small-world condition does not reproduce the paper's decrease in global disagreement, and the random-graph condition is milder than the paper on polarization and NCI magnitude. We interpret these differences cautiously. The paper reports point estimates from stochastic LLM trajectories, while our runs average five current-API trajectories. Because LLM outputs, parsing, and API-side behavior can vary across time, exact trajectory-level agreement is unlikely. The important point for the framework claim is that the core trends remain: scale-free and small-world graphs produce rising polarization and local alignment, while the random graph does not produce the same coherent disagreement-reduction pattern as the scale-free graph. Importantly, we run simulations from the original paper's codebase, which similarly diverge from the paper's results while maintaining qualitative trends.

\textbf{Interpretation.} H1 supports the claim that EchoSim's assumptions can be decomposed into EASE components without destroying the main phenomenon. The graph topology belongs to the environment, similarity exposure belongs to observation formation, short/long memory belongs to the agent, repeated social days belong to the engine, and echo-chamber indices belong to evaluation. This modular mapping is what enables the following ablations.

\paragraph{H2. Opposing recommendation timelines should weaken polarization.}
\label{app:opposing}
Having reproduced the baseline, we next vary only the observation-formation policy. The original setup exposes agents to belief-similar neighbors. We replace this with opposing exposure, where agents preferentially observe neighbors whose beliefs differ substantially.

\textbf{What was varied.} Recommendation/observation policy on the same fixed scale-free graph:
\begin{enumerate}
    \item Similarity exposure: agents observe neighbors with similar beliefs.
    \item Opposing exposure: agents observe neighbors with distant or opposing beliefs.
    \item Random exposure: agents observe eligible neighbors without belief-similarity filtering.
\end{enumerate}

\textbf{Outcome.} Opposing exposure strongly reduces final polarization and global disagreement relative to similarity exposure. Final polarization drops from $2.722$ to $1.796$, and final global disagreement drops from $2.228$ to $1.557$. Random exposure produces a weaker version of the same moderation pattern, reducing final polarization to $2.431$. However, neither opposing nor random exposure reduces final NCI: opposing exposure ends at $0.412$ and random exposure at $0.447$, compared with $0.382$ under similarity.

\textbf{Interpretation.} H2 is partially supported. Changing exposure away from similarity reduces population-level polarization, which is the expected direction if agents encounter a broader belief range. However, NCI does not fall. One plausible explanation is that exposure to opposing views does not necessarily mix local neighborhoods; instead, it can move groups internally or create new local consistency after belief updates. In other words, global polarization and local sorting are separable. This is useful for the paper's argument: because EASE isolates observation formation, we can see that changing the recommender changes some macro-metrics while leaving others unexpectedly intact.

\paragraph{H3. Echo-chamber behavior persists after loosening the action structure.}
The original simulator has a strict opinion-dynamics structure: agents receive selected neighbor opinions, summarize short-term memory, consolidate long-term memory, and update belief once per day. This is clean, but it assumes that social media interaction is essentially direct opinion exposure. We loosen this by replacing the direct opinion interface with a Twitter-like environment where agents can post, repost, reply, or like. Belief is measured through a structured end-of-window probe.

\textbf{What was varied.} Environment-agent interaction layer:
\begin{enumerate}
    \item Exact reproduction: direct neighbor opinion exposure and daily belief update.
    \item Loose social environment: timeline observations, social-media actions, and terminal belief probe.
\end{enumerate}

\textbf{Outcome.} The loose social environment still produces the qualitative echo-chamber signature. With Echo-style memory and self-state feedback, final polarization reaches $2.990\pm0.150$, NCI reaches $0.411\pm0.108$, and global disagreement falls to $2.296\pm0.085$. From the shared initial state, this corresponds to $\Delta$polarization $=+0.858$, $\Delta$NCI $=+0.608$, and $\Delta$global disagreement $=-0.295$.

\textbf{Interpretation.} H3 suggests that the original result is not merely an artifact of the strict direct-opinion action interface. Even when agents act through a more realistic social-media surface, the combination of graph structure, timeline exposure, and belief updating still produces polarization and local alignment. Mechanistically, the social app changes how information is expressed, but it does not remove repeated exposure to socially local content. The result therefore strengthens the claim that the phenomenon is robust to a change in environment-agent interface.

\paragraph{H4. Self-state feedback anchors belief trajectories.}
EchoSim agents are explicitly reminded of their current opinion and belief during belief updating. This design choice is easy to treat as bookkeeping, but prior work suggests that feeding agents their previously reported states can affect belief consistency \citep{bückkaeffer2026textitsiliconsocietycookbookdesign}. We therefore test whether self-state feedback is a substantive agent assumption.

\textbf{What was varied.} Belief-update context:
\begin{enumerate}
    \item With self-state feedback: the agent is reminded of its previous opinion and belief.
    \item Without self-state feedback: the same observations and memory are provided, but explicit previous opinion/belief fields are removed.
\end{enumerate}

\textbf{Outcome.} Removing self-state feedback weakens polarization and local alignment. Under GPT-4o-mini with Echo-style memory, final polarization falls from $2.990$ to $2.695$, and final NCI falls from $0.411$ to $0.295$. Belief volatility increases from $0.100$ to $0.238$ mean absolute belief change per agent-step.

\textbf{Interpretation.} H4 is supported. Self-state feedback appears to act as an anchoring mechanism. When agents are reminded of their own previously reported belief, they produce more stable and more polarized trajectories. When this reminder is removed, agents remain exposed to the same social environment but become more volatile and less locally aligned. This shows why EASE treats prompt/state construction as an agent component rather than an implementation detail.

\paragraph{H5. Simpler memory agents preserve weaker echo-chamber behavior.}
We next vary the agent cognitive architecture. EchoSim includes explicit short-term memory summarization and long-term memory consolidation. We replace this with a simpler observe-memory-act social agent that receives observations and answers the same structured belief probe, but does not reproduce the explicit EchoSim short/long-memory pipeline.

\textbf{What was varied.} Agent memory architecture:
\begin{enumerate}
    \item Echo-memory agent: preserves short-term summary, long-term consolidation, and structured belief update.
    \item Simple social agent: uses a simpler observe-memory-act memory path.
\end{enumerate}

\textbf{Outcome.} The simple social agent still shows echo-chamber directionality, but the effect is weaker. With self-state feedback, final polarization falls from $2.990$ for the Echo-memory agent to $2.641$ for the simple agent, and NCI falls from $0.411$ to $0.193$. Without self-state feedback, the simple agent weakens further, ending at polarization $2.391$ and NCI $0.120$.

\textbf{Interpretation.} H5 is supported in a graded sense. The echo-chamber pattern does not require the full EchoSim memory architecture, but that architecture amplifies the phenomenon. A likely mechanism is that short/long-memory consolidation compresses repeated exposures into a coherent self-narrative, making later belief reports more consistent and more extreme. The simple agent sees similar social content but does less structured consolidation, so it still moves, but with weaker local sorting.

\paragraph{H6. Model backbone changes the echo-chamber signature.}
Finally, we vary the LLM backbone by replacing GPT-4o-mini with Qwen3.5-4B. This tests whether the echo-chamber result is mainly a property of the environment and graph, or whether it depends on model-level behavior.

\textbf{What was varied.} LLM backbone:
\begin{enumerate}
    \item GPT-4o-mini.
    \item Qwen3.5-4B.
\end{enumerate}

\textbf{Outcome.} Qwen3.5-4B does not reproduce the GPT-like local-alignment signature. With Echo memory and self-state feedback, Qwen reaches final polarization $2.605$, but NCI remains negative at $-0.110$, and global disagreement increases to $2.841$. Without self-state feedback, Qwen Echo-memory agents become highly volatile, with mean belief volatility $0.501$. Qwen simple agents are more stable, but show only weak polarization growth; the no-self-state simple condition rises only from $2.132$ to $2.298$ in polarization.

\begin{table}[t]
\centering
\small
\begin{tabular}{lrrrr}
\toprule
\textbf{Regime} & \textbf{Polarization} & \textbf{NCI} & \textbf{Global Disagreement} & \textbf{Volatility} \\
\midrule
GPT Echo + self-state & $2.990\pm0.150$ & $0.411\pm0.108$ & $2.296\pm0.085$ & $0.100$ \\
GPT Echo no self-state & $2.695\pm0.112$ & $0.295\pm0.059$ & $2.273\pm0.089$ & $0.238$ \\
GPT simple + self-state & $2.641\pm0.143$ & $0.193\pm0.020$ & $2.379\pm0.117$ & $0.124$ \\
GPT simple no self-state & $2.391\pm0.065$ & $0.120\pm0.059$ & $2.317\pm0.117$ & $0.195$ \\
\midrule
Qwen Echo + self-state & $2.605\pm0.249$ & $-0.110\pm0.143$ & $2.841\pm0.378$ & $0.200$ \\
Qwen Echo no self-state & $2.485\pm0.146$ & $-0.114\pm0.084$ & $2.692\pm0.186$ & $0.501$ \\
Qwen simple + self-state & $2.437\pm0.122$ & $0.034\pm0.098$ & $2.439\pm0.174$ & $0.052$ \\
Qwen simple no self-state & $2.298\pm0.108$ & $0.033\pm0.121$ & $2.331\pm0.147$ & $0.070$ \\
\bottomrule
\end{tabular}
\caption{Final loose-social metrics across model, memory architecture, and self-state feedback conditions. Volatility is mean absolute belief change per agent-step.}
\label{tab:echo-loose-social}
\end{table}

\textbf{Interpretation.} H6 shows that the LLM backbone is not a neutral plug-in. GPT-4o-mini produces the expected combination of polarization, local alignment, and reduced neighbor disagreement. Qwen3.5-4B produces belief movement, but not the same neighborhood-correlation structure. The interaction with memory is especially informative: Qwen Echo-memory agents are volatile, while Qwen simple agents are stable but weakly polarizing. This suggests that memory consolidation can amplify model-specific tendencies rather than uniformly improving simulation fidelity.

\paragraph{Discussion.}
This case study demonstrates how EASE turns reproduction into component-wise simulator analysis. Additionally, a monolithic reproduction could only ask whether EchoSim replicates. A more thorough analysis would require a massive ablation study over components. The study schema provides a more efficient way to wade through the component space by generating and validating hypotheses. The initial reproduction shows that the original scale-free echo-chamber pattern survives decomposition into EASE. The exposure-policy ablation shows that observation formation can reduce polarization without simply reducing local alignment. The loose-social experiment shows that the phenomenon is not confined to direct opinion exchange. The self-state and memory ablations show that agent cognition and prompt-state construction materially shape polarization and volatility. Finally, the Qwen comparison shows that the LLM backbone changes not just magnitude, but the qualitative metric signature.

Together, these results support the broader paper's claim: LLM social simulations should be modular and configurable because aggregate outcomes are produced by interacting assumptions across the agent, environment, simulation engine, and evaluation layers. EASE makes those assumptions explicit enough to reproduce, relax, and falsify one component at a time.

\section{Study descriptions}\label{sec:studies}
\paragraph{Style Diversity} We performed a style diversity study with the following filled-out schema:
\begin{verbatim}
study:
  name: style_diversity
  question: Does increasing LLM capacity reduce repetitive/groupthink behavior in
    multi-agent social media simulations?
  scenarios:
  - ai_conference
  - misinformation
  run_overrides:
    simulation.execution.max_steps: 10
hypotheses:
  h1_model_capacity:
    statement: Larger language models produce more diverse agent behavior (higher
      lexical diversity, lower self-BLEU, more varied actions).
    independent_variable: model
    prediction: gpt4o outperforms gpt4o-mini on diversity metrics across scenarios.
    status: testing
    conditions:
      gpt4o-mini:
        cli_override: model=gpt4omini
        runs:
        - scenario: ai_conference
          source: outputs/ai_conference_experiment/2026-03-29_20-15-50
          eval: outputs/eval_style_diversity/h1_model_capacity/gpt4o-mini/ai_conference/eval.json
        - scenario: misinformation
          source: outputs/misinformation_experiment/2026-03-29_20-31-36
          eval: outputs/eval_style_diversity/h1_model_capacity/gpt4o-mini/misinformation/eval.json
      gpt4o:
        cli_override: model=gpt4o
        runs:
        - scenario: ai_conference
          source: outputs/ai_conference_experiment/2026-03-29_20-37-13
          eval: outputs/eval_style_diversity/h1_model_capacity/gpt4o/ai_conference/eval.json
        - scenario: misinformation
          source: outputs/misinformation_experiment/2026-03-29_20-43-00
          eval: outputs/eval_style_diversity/h1_model_capacity/gpt4o/misinformation/eval.json
analysis:
  comparison: outputs/eval_style_diversity/h1_model_capacity/comparison/eval_comparison.json
\end{verbatim}

\section{Study Details}

\subsection{Intervention Prompts for Engagement Case Study}
\label{app:intervention_prompts}

\begin{tcolorbox}[
    colback=gray!5!white,      
    colframe=gray!75!black,    
    colbacktitle=gray!20!white,
    coltitle=black,            
    fonttitle=\bfseries,       
    title={Intervention: Mild}, 
    boxrule=1pt,               
    arc=3pt,                   
    left=8pt, right=8pt, top=8pt, bottom=8pt 
]
\textbf{IMPORTANT:} Act realistically following your given persona: Consider what actions, and how many of them the persona you are simulating would take given the timeline observations in the current session. If continuing a session, also consider the actions you have already taken in the step.
\end{tcolorbox}

\begin{tcolorbox}[
    colback=gray!5!white,      
    colframe=gray!75!black,    
    colbacktitle=gray!20!white,
    coltitle=black,            
    fonttitle=\bfseries,       
    title={Intervention: Moderate}, 
    boxrule=1pt,               
    arc=3pt,                   
    left=8pt, right=8pt, top=8pt, bottom=8pt 
]
\textbf{IMPORTANT:} Act realistically following your given persona: Consider what actions, and how many of them the persona you are simulating would take given the timeline observations in the current session. Remember that humans typically take a very limited number of actions on average, so strictly consider your persona and past behavior when deciding actions. If continuing a session, also consider the actions you have already taken in the step.
\end{tcolorbox}

\begin{tcolorbox}[
    colback=gray!5!white,      
    colframe=gray!75!black,    
    colbacktitle=gray!20!white,
    coltitle=black,            
    fonttitle=\bfseries,       
    title={Intervention: Strict}, 
    boxrule=1pt,               
    arc=3pt,                   
    left=8pt, right=8pt, top=8pt, bottom=8pt 
]
\textbf{IMPORTANT:} Act realistically following your given persona: Consider what actions, and how many of them the persona you are simulating would take given the timeline observations in the current session. Remember that humans typically take a very limited number of actions on average, so strictly consider your persona and past behavior when deciding actions. If continuing a session, also consider the actions you have already taken in the step, and ensure the total number of actions given the timeline posts is realistic. Take the finished step instead of over-interacting/posting when you believe the agent's key desired actions have been conducted.
\end{tcolorbox}

\subsection{Further Hypothesis for Engagement Dynamics}
\label{app:hyp}
\begin{enumerate}
    \item \textbf{H5}: Algorithmic recommendation system feeds lead to more realistic information dynamic structures such as cascade measurements, virality etc.
    \item \textbf{H6}: Agents can be aligned to real-world distributions for engagement-actions via agent selection or assigning agents pre-set social personas.
    \item \textbf{H7}: By explicitly making the follower-chronological field non-interesting to the user (not aligned with voting goal, or interests), we can elicit a much bigger gap between the recsys-TWHiN and the chronological timeline
\end{enumerate}

\newpage

\end{document}